\documentclass[%
 reprint,showpacs,showkeys,
 amsmath,amssymb,
 aps,
]{revtex4-1}

\usepackage{graphicx}
\usepackage{epstopdf}
\usepackage{dcolumn}
\usepackage{bm}
\usepackage{hyperref} 

\begin{document}

\preprint{APS/123-QED}

\title{Dynamics of Coulombic and Gravitational Periodic Systems}
\author{Pankaj Kumar}
\author{Bruce N. Miller}%
 \email{b.miller@tcu.edu}

\affiliation{
 Department of Physics and Astronomy,\\
 Texas Christian University, Fort Worth, TX 76129.
}

\date{\today}

\begin{abstract}
We study the dynamics and the phase-space structures of Coulombic and self-gravitating versions of the classical one-dimensional 3-body system with periodic boundary conditions. We demonstrate that such a 3-body system may be reduced isomorphically to a spatially periodic system of a single particle experiencing a two-dimensional potential on a rhombic plane. For the case of both Coulombic and gravitational versions, exact expressions of the Hamiltonian have been derived in rhombic coordinates. We simulate the phase-space evolution through an event-driven algorithm that utilizes analytic solutions to the equations of motion. The simulation results show that the motion exhibits chaotic, quasiperiodic, and periodic behaviors in segmented regions of the phase space. While there is no evidence of global chaos in either the Coulombic or the gravitational system, the former exhibits a transition from a completely non-chaotic phase space at low energies to a mixed behavior. Gradual yet striking transitions from mild to intense chaos are indicated with changing energy, a behavior that differentiates the spatially periodic systems studied in this paper from the well-understood free-boundary versions of the 3-body problem. Our treatment of the 3-body systems is the first one of its kind and opens avenues for analysis of the dynamical properties exhibited by spatially periodic versions of various classes of systems
studied in plasma and gravitational physics as well as in cosmology.
\end{abstract}

\maketitle


\begin{figure*}[t]
\includegraphics[scale=0.62]{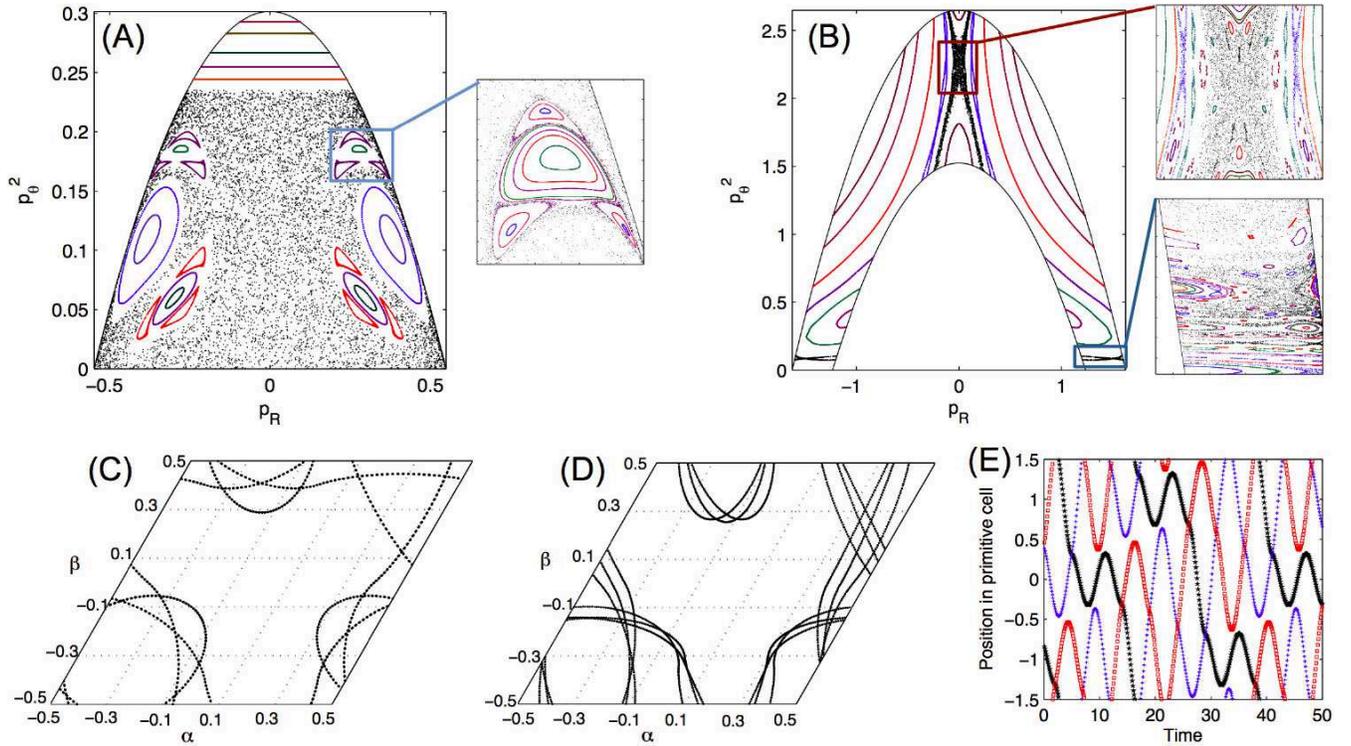}
\caption{\label{fig:Plasma} Poincar\'e plots for the Coulombic system with (A) ${\mathcal{H}}_c = -1.098$ and (B) ${\mathcal{H}}_c = 2.031$. The boxes denote the areas magnified in the corresponding insets. (C) and (D) depict the periodic trajectories on the rhombic plane for a P2 orbit and a P6 orbit respectively, each with ${\mathcal{H}}_c = -1.098$. (E) Position vs time for the three-particle system corresponding to the rhombic trajectory shown in (C). $\alpha$ and $\beta$ have been expressed in the units of $\sqrt{2}L$.}
\end{figure*}
\begin{figure*}[t]
\includegraphics[scale=0.78]{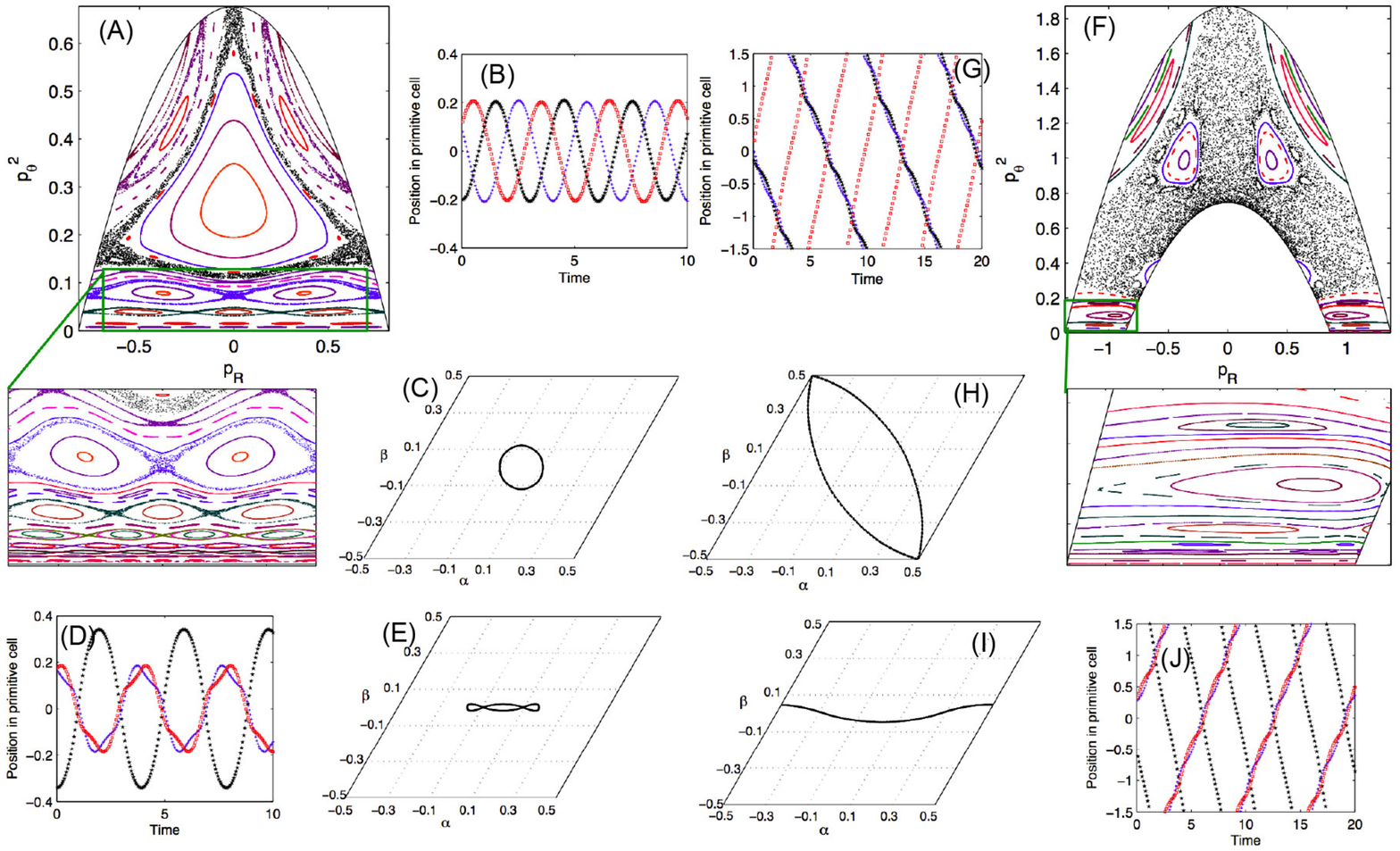}
\caption{\label{fig:Gravity} Simulation results for the gravitating system: (A-E) are for ${\mathcal{H}}_g = 0.226$ whereas (F-H) are for ${\mathcal{H}}_g = 0.624$. (A) and (F) show Poincar\'e plots for the two values of ${\mathcal{H}}_g$. The boxes denote the areas magnified in the corresponding insets.  The three-particle evolution and the corresponding trajectories on the rhombic plane for: a P1 orbit (B-C); a P3 orbit (D-E); a P2 orbit (G-H); another P2 orbit (I-J). $\alpha$ and $\beta$ have been expressed in the units of $\sqrt{2}L$.}
\end{figure*}

Physicists often rely on one-dimensional models as a starting point in the analysis of the more complicated higher dimensional systems. Not only can they be mapped to systems subject to experimental study \cite{Milner2001}, a one-dimensional gravitation-like interaction has actually been observed in the laboratory \cite{Chalony2013}. Moreover, one-dimensional systems have been of great intrinsic interest to theoretical and computational physicists (see \cite{Rybicki1971,Miller1997,Miller1998,Lauritzen2013,Kumar2014} and references therein). Small versions of the one-dimensional systems ($N\geq 3$) are of particular interest as they are relatively less convoluted to analyze yet may exhibit interesting chaotic dynamics.

The simplest nontrivial case is the 3-body problem. While classical \cite{Lehtihet1986} as well as relativistic \cite{Burnell2003,Burnell2004} and post-Newtonian \cite{Burnell2004} versions of one-dimensional three-body problems with free boundary conditions have been studied in considerable detail, such treatments have not been extended to the periodic versions of the system. In such studies in the field of plasma physics and cosmology, periodic boundary conditions have long been preferred  \cite{Springiel2006,Bertschinger1998,hockney1988,Hernquist1991} and have been employed in one-dimensional Coulombic and gravitational models \cite{Kunz1974,Schotte1980,miller2010,Kumar2014}. The question at hand is: how do we reduce a three-body one-dimensional problem with periodic boundary conditions to one of a single-body in two dimensions to study the dynamical properties of the system? Here we present the results of the first study (to the best of our knowledge) of chaos in periodic versions of the 3-body problem.

We consider two versions of a spatially periodic lineal system on an $x$-axis with the primitive cell extending in $[-L,L)$ and which contains three infinite sheets, each with a surface mass density $m$, located at $x_1$, $x_2$, and $x_3$ with respect to the center of the cell. In one version, the sheets are uncharged and are only interacting gravitationally. The other version is essentially a one-dimensional Coulombic system in which the sheets are charged with a surface charge density $q$ and are immersed in a uniformly distributed negative background such that the net charge is zero. For the case of the charged version of the system, we neglect the gravitational effects and take into account only the Coulomb interactions. Using results from \cite{miller2010,Kumar2014}, the potential energy for the system may be expressed as
\begin{equation} \label{eq_potential_energy3P}
\mathcal{V} = \kappa \sum_{i<j}^{3} \left(\frac{(x_j-x_i)^2}{2L} - |x_j-x_i|\right),
\end{equation}
where $\kappa = \frac{2\pi kq^2}{m}$ for the case of the Coulombic system (with each sheet, henceforth referred to as a particle or a body, having a surface charge density $q$ in addition to the surface mass density $m$) and $\kappa = -2\pi Gm^2$ for the gravitating system. As it is common practice, we henceforth refer to the sheets as particles or bodies.

Without losing generality, we set the initial center-of-mass velocity to be zero. Then, with one constraint set forth by conservation of momentum on each of the configuration and the momentum subspaces, the phase-space dimensionality may be reduced by two. Normally, for a one-dimensional 3-body system, this reduction is often accomplished by choosing a two-dimensional orthornormal basis, say $\{\alpha,\beta\}$, in the configuration space which invokes a two-dimensional orthornormal conjugate basis (oftentimes simply the corresponding velocities) in the momentum space. This way, a 3-body one-dimensional problem is reduced to one with a single-body on a plane \cite{Lehtihet1986}. However, the systems we have at hand are spatially periodic and hence the reduced system must also appropriately exhibit spatial periodicity.

In dealing with the problem, one realizes that moving away from an otherwise traditional orthornormal basis is perhaps the only way to democratically realize periodicity in the reduced system. In other words, we look for a set of coordinates that inherently embodies the periodicity of the system. Notice that the particles in our system may be thought of being on a ring (or a torus) of circumference $2L$ \cite{miller2010,Kumar2014}. In this torus representation, we realize that even though the separations in the primitive cell, $(x_j-x_i)$, may assume any value in the interval $[-2L,2L)$, the separations on the torus may only be in $[-L,L)$. In other words, relative separations on the torus naturally inherit the spatial periodicity, we choose $\alpha$ and $\beta$ to be the normalized relative separations of the particle at $x_2$ with respect to the one at $x_1$ and the particle at $x_3$ with respect to the one at $x_2$ respectively on the torus. That is,
\begin{equation} \label{eq_alpha_coordinates_3p}
\sqrt{2}\alpha = \left\lbrace
\begin{array}{c r}
\left. (x_2-x_1) + 2L \right., &  -2L \leq (x_2-x_1) < -L \\
\left. (x_2-x_1) \right., &  -L \leq (x_2-x_1) < L \\
\left. (x_2-x_1) - 2L \right., &  L \leq (x_2-x_1) < 2L \\
\end{array}\right.\,.
\end{equation}

Similarly, we define the second spatial coordinate, $\beta$, by simply replacing $(x_2-x_1)$ by $(x_3-x_2)$ in Eq. (\ref{eq_alpha_coordinates_3p}). Note that while $\alpha$ and $\beta$ undergo discontinuity, their corresponding conjugate momenta, $p_{\alpha}$ and $p_{\beta}$, do not. This is because the three individual momenta ($p_1$, $p_2$, and $p_3$) in the original spatially-periodic Coulomb and gravitating systems are continuous \cite{miller2010,Kumar2014} and an addition (or a subtraction) of the constant quantity $2L$ does not affect the time derivatives of the relative separations. Therefore, we define the two momenta as
\begin{equation} \label{eq_alpha_dot_coordinates_3p}
\sqrt{2}p_{\alpha} = (p_2 - p_1), \hspace{15pt} \sqrt{2}p_{\beta} = (p_3 - p_2).
\end{equation}
With the above transformation of coordinates, the Hamiltonian may be expressed as
\begin{eqnarray} \label{eq_Hamiltonian_alpha_beta_3P}
\mathcal{H}=\frac{2}{3m} \left({p_{\alpha}}^2 + {p_{\beta}}^2 + {p_{\alpha}}{p_{\beta}} \right) +\kappa \left[\frac{\alpha^2 + \beta^2}{L} + \right. \nonumber \\
\left.\frac{(\alpha + \beta)^2}{L} - \sqrt{2}\left(|\alpha| + |\beta| + |\alpha + \beta|\right)\right].
\end{eqnarray}
Clearly, in the new system of coordinates, the kinetic energy, $\mathcal{T} = \frac{2}{3}m({p_{\alpha}}^2 + {p_{\beta}}^2 + {p_{\alpha}}{p_{\beta}})$, is non-diagonal. In fact, the form that $\mathcal{T}(p_{\alpha},p_{\beta})$ takes suggests that Eq. (\ref{eq_Hamiltonian_alpha_beta_3P}) represents the Hamiltonian of a single particle expressed in rhombic coordinates, with the two configuration (as well as conjugate-momentum) coordinates inclined at an angle of $60^o$. In other words, by invoking the transformations as defined in Eqs. (\ref{eq_alpha_coordinates_3p}) and (\ref{eq_alpha_dot_coordinates_3p}), we have reduced the dynamics of the one-dimensional, spatially-periodic, three-particle system to that of a two-dimensional periodic system of a single particle on a $60^o$-rhombic plane of edge length $\sqrt{2}L$.

Time evolution of $\alpha$ and $\beta$ is governed by Hamilton's equations. For example, the equations of motion for, say, the coordinate $\alpha$ may be obtained from
\begin{equation}
\frac{d}{dt}\left(\frac{\partial }{\partial p_{\alpha}} \mathcal{T}({p_{\alpha}},{p_{\beta}})\right) = -\frac{\partial}{\partial \alpha}\mathcal{V}(\alpha, \beta),
\end{equation}
where $\mathcal{V}(\alpha, \beta)$ is the contribution of the potential energy to the Hamiltonian in Eq. (\ref{eq_Hamiltonian_alpha_beta_3P}). With an expression obtained similarly for the coordinate $\beta$, it can be shown that the resulting set of coupled differential equations is analytically solvable between the events of interparticle crossings, which on the rhombic plane correspond to one of the following three conditions:
\begin{equation} \label{eq_crossing_conditions}
\alpha=0, \hspace{10pt} \beta=0, \hspace{10pt} (\alpha+\beta)=0.
\end{equation}
We track each of the above conditions in simulation through an event-driven algorithm based on the approach adopted by Kumar and Miller \cite{Kumar2014}. The algorithm utilizes exact mathematical equations to determine the positions and velocities for each iteration. With the ability to follow the time evolution, we plot Poincar\'e surfaces of section for the Coulombic system and gravitating systems to study the global structures of their phase spaces. For consistency and to facilitate easy comparison of our results with those obtained earlier \cite{Lehtihet1986,Burnell2003}, the Poincar\'e surfaces have been generated by plotting the square of the angular component of the momentum (${p_{\theta}}^2$) versus the radial component of the momentum ($p_R$) for $\beta = 0$, that is, for the events of particle 2 undergoing a crossing with particle 3 in the three-particle system. It should be noted that while the potential energy in the free-boundary versions of the three-particle self-gravitating system could attain any non-negative value, the potential energy in the spatially-periodic gravitational as well as Coulombic systems are bounded \cite{miller2010,Kumar2014}. This sets constraints on the values that $p_R$ and $p^2_{\theta}$ may attain for any given value of the Hamiltonian. With $\beta = 0$, it can be shown rigorously that $p_R$ and $p_{\theta}$ satisfy
\begin{equation}
2m'{\mathcal{H}}_c - {p_R}^2 \leq {p_{\theta}}^2 \leq 2m'({\mathcal{H}}_c + 2\pi kq^2L) - {p_R}^2
\end{equation}
for the Coulombic system whereas for the gravitational system, they satisfy
\begin{equation}
2m'({\mathcal{H}}_g-2\pi Gm^2L) - {p_R}^2 \leq {p_{\theta}}^2 \leq 2m'{\mathcal{H}}_g - {p_R}^2,
\end{equation}
where $m' \equiv (4/3)m$, and the subscripts $c$ and $g$ in the Hamiltonian, $\mathcal{H}$ represent the Coulombic and gravitational systems respectively.

We perform the simulations in dimensionless units as proposed for Coulombic and gravitational systems in \cite{miller2010} and \cite{Kumar2014} respectively. For easier comparison between the two versions, dimensionless momenta, $p_R$ and $p_{\theta}$ have further been expressed in terms of $m'$. We have presented simulation results for a low-energy and a high-energy configuration each for the Coulombic as well the gravitating system. In each case, we plot the Poincar\'e surfaces with primitive-cell evolution of the three-particle system and their corresponding trajectories on the rhombic plane for a few periodic orbits of interest. On the Poincar\'e surfaces, chaotic orbits appear as filled-in regions, quasiperiodic orbits take the form of continuous or segmented curves whereas purely periodic orbits show up as finite number of isolated dots (usually surrounded by quasiperiodic orbits). (Note for readers of the online version: Colors represent different initial conditions).

Note that, in the Coulomb system, no interparticle crossing takes place for ${\mathcal{H}}_c<-1.500$, and hence, Poincar\'e surfaces satisfying $\beta = 0$ may not be plotted. The particles simply undergo pure oscillations about the equilibrium points for ${\mathcal{H}}_c<-1.500$, and therefore, the motion is strictly periodic. For $-1.500<{\mathcal{H}}_c<-1.350$, the entire Poincar\'e plot is characterized by parallel horizontal ``lines'' similar to the topmost ``horizontal'' section of Fig. \ref{fig:Plasma}(A). Such behavior, which is also seen in the wedge billiard at a half angle of $45^\circ$ \cite{Lehtihet1986}, suggests integrability. However, as the energy is increased to ${\mathcal{H}}_c\sim-1.350$, a narrow chaotic region appears at the bottom growing upward,  resulting in the reduction of the integrable segment. This is in contrast to the wedge billiard at $45^\circ$ in which the complete phase space is integrable whereas, in the periodic Coulombic system, it appears that both integrable and chaotic regions may coexist in the phase space depending on the energy. To our knowledge this is the first observation of such behavior.

As the energy is further increased, stable islands appear in the chaotic part through what appear to be period-doubling bifurcations. In Fig. \ref{fig:Plasma}(A), which was generated for ${\mathcal{H}}_c=-1.098$, each of the``closed loops'' converge to a single periodic point. Figure \ref{fig:Plasma}(C) shows the periodic trajectory on the rhombus corresponding to the centers of the two large stable regions on the sides. We refer to such an orbit as a``period-2'' orbit or simply a P2 orbit. Figure \ref{fig:Plasma}(D) shows the periodic trajectory corresponding to the P6 orbit found in each of the lobes in the boxed region on Fig. \ref{fig:Plasma}(A) (and its mirror image on the left). The primitive-cell evolution corresponding to the trajectory in Fig. \ref{fig:Plasma}(C) has been shown in Fig. \ref{fig:Plasma}(E).

With further increase in the energy of the Coulombic system, the integrable ``horizontal'' region is squeezed upward to eventually disappear. At the same time, the stable islands within the chaotic region grow progressively in size taking over most of the Poincar\'e surface, and leading to the plot presented in Fig \ref{fig:Plasma}(B) at ${\mathcal{H}}_c=2.031$. Notice that the stable regions have grown large enough to merge, sectioning the chaotic region into three a narrow regions,  one in the upper-central region and two near tips of the ``horseshoe.'' When the plot is enlarged, we can see fine structures of stable and unstable regions forming within those narrow chaotic regions.

While the Coulombic system shows no chaos at low energies, the gravitating counterpart tells a completely different story. At low energies, the behavior resembles that of the three-body gravitating system with free (also known as open) boundary conditions. This suggests that at low-energies, each primitive-cell of the periodic gravitational system can be mapped to the physical systems considered in \cite{Lehtihet1986,Milner2001}. At first, this observation appears rather surprising. However, if we look closely at the potential, we realize that at low energies, the interparticle separations are very small, and therefore, the quadratic terms in the potential may be disregarded. The low-energy Poincar\'e plot has been discussed in \cite{Lehtihet1986,Burnell2003}.

If we compare the results for ${\mathcal{H}}_g=0.226$ shown in Fig. \ref{fig:Gravity}(A) with those presented for the free-boundary versions of the self-gravitating system \cite{Lehtihet1986,Burnell2003}, we see striking similarities. Three major stable regions form the central and the upper left and right portions of the plot. The fractal region, located in the lower part of the plot, exhibits self-similar sets of nested stable islands and includes infinite ``period-N'' orbits that are surrounded by quasiperiodic orbits between which narrow chaotic regions exist. Figure \ref{fig:Gravity}(B) shows the primitive cell evolution for the P1 orbit from \ref{fig:Gravity} (A) whereas \ref{fig:Gravity} (C) shows the corresponding motion on the rhombic plane, with Figs. \ref{fig:Gravity}(D) and (E) showing the same for the P3 orbit from the fractal part of the Poincar\'e plot. Notice that Fig. \ref{fig:Gravity}(E) is nothing but a pretzel orbit as discussed in \cite{Burnell2003}.

As energy is increased, the gravitating system starts deviating from the free-boundary behavior. Note that, in the classical free-boundary version of the gravitating system, changing energy does not bring about a change in the structure of the phase-space, it just scales the phase space. However, in the periodic version, as energy is increased from a low value, the phase space gets more chaotic. However, small stable islands start appearing, as seen in Fig. \ref{fig:Gravity}(F) for ${\mathcal{H}}_g=0.624$, which grow and finally engulf the chaotic region. At higher energies, the particles are able to cross between rhombic planes (or equivalently, adjacent cells in the periodic 3-body system). Plots of primitive cell evolution and motion on the rhombic plane are shown for the central P2 orbit in Fig. \ref{fig:Gravity}(F) in Figs. \ref{fig:Gravity}(G) and (H) respectively. Another set is shown for a different P2 orbit (lying at the bottom of the horseshoe) in Figs. \ref{fig:Gravity}(I) and (J) respectively. Also note that periodic orbits form ``closed loops'' (for the periodic boundary conditions, it means that after a finite number of strands on the rhombic plane, the trajectory will simply repeat on top of each other), quasiperiodic orbits result in ``bands'' on the rhombic plane that trajectories will never move out of, and chaotic orbits lead to unpredictable trajectories on the plane. Of course, conservation of energy may set a boundary beyond which the particle will never go, in which case, the chaotic trajectory will fill in the allowed region as time progresses). As one moves away from a fixed point (a ``period N'' trajectory, N $\in I$) on the Poincar\'e plot, quasiperiodicity takes over and the strands on the rhombus starts spreading into bands.

Notice that for both the Coulombic and gravitational systems, the Poincar\'e horseshoe itself gets narrower with increasing energy.
While we have compared the results of gravitating system with those already known for the different free-boundary cases, and we have compared the Coulombic system with the gravitational version, the results for Coulombic system still stand out and are of particular interest because this is the first analysis, as far as we know, of the phase-space structure for a purely Coulombic system. However, in the periodic versions of both Coulombic and self-gravitating systems, we do confirm the presence of considerable structure which prevents KAM breakdown to global chaos \cite{Arnold1968}.

The 3-body spatially periodic systems provide the first theoretical basis for understanding the chaotic dynamics of small systems with periodic boundary conditions. This study reveals striking similarities and differences compared with their free-boundary counterparts not previously seen and points toward the need for studying the 3-body periodic Colombic and gravitational systems in greater detail. Furthermore, analytic reduction of the systems' equations of motion in rhombic coordinates may be utilized to extend the tangent-space approach \cite{Benettin1979} for calculating the spectrum of Lyapunov exponents for these spatially periodic systems in simulation. Our work also paves the way to study the dynamics of 4-body periodic versions, the results of which could be compared with those already known for the 4-body free-boundary gravitational case \cite{Lauritzen2013}.

\bibliography{KumarMiller_ChaoticDynamicsVer1}

\providecommand{\noopsort}[1]{}\providecommand{\singleletter}[1]{#1}%
\begin{thebibliography}{19}%
\makeatletter
\providecommand \@ifxundefined [1]{%
 \@ifx{#1\undefined}
}%
\providecommand \@ifnum [1]{%
 \ifnum #1\expandafter \@firstoftwo
 \else \expandafter \@secondoftwo
 \fi
}%
\providecommand \@ifx [1]{%
 \ifx #1\expandafter \@firstoftwo
 \else \expandafter \@secondoftwo
 \fi
}%
\providecommand \natexlab [1]{#1}%
\providecommand \enquote  [1]{``#1''}%
\providecommand \bibnamefont  [1]{#1}%
\providecommand \bibfnamefont [1]{#1}%
\providecommand \citenamefont [1]{#1}%
\providecommand \href@noop [0]{\@secondoftwo}%
\providecommand \href [0]{\begingroup \@sanitize@url \@href}%
\providecommand \@href[1]{\@@startlink{#1}\@@href}%
\providecommand \@@href[1]{\endgroup#1\@@endlink}%
\providecommand \@sanitize@url [0]{\catcode `\\12\catcode `\$12\catcode
  `\&12\catcode `\#12\catcode `\^12\catcode `\_12\catcode `\%12\relax}%
\providecommand \@@startlink[1]{}%
\providecommand \@@endlink[0]{}%
\providecommand \url  [0]{\begingroup\@sanitize@url \@url }%
\providecommand \@url [1]{\endgroup\@href {#1}{\urlprefix }}%
\providecommand \urlprefix  [0]{URL }%
\providecommand \Eprint [0]{\href }%
\providecommand \doibase [0]{http://dx.doi.org/}%
\providecommand \selectlanguage [0]{\@gobble}%
\providecommand \bibinfo  [0]{\@secondoftwo}%
\providecommand \bibfield  [0]{\@secondoftwo}%
\providecommand \translation [1]{[#1]}%
\providecommand \BibitemOpen [0]{}%
\providecommand \bibitemStop [0]{}%
\providecommand \bibitemNoStop [0]{.\EOS\space}%
\providecommand \EOS [0]{\spacefactor3000\relax}%
\providecommand \BibitemShut  [1]{\csname bibitem#1\endcsname}%
\let\auto@bib@innerbib\@empty
\bibitem [{\citenamefont {Milner}\ \emph {et~al.}(2001)\citenamefont {Milner},
  \citenamefont {Hanssen}, \citenamefont {Campbell},\ and\ \citenamefont
  {Raizen}}]{Milner2001}%
  \BibitemOpen
  \bibfield  {author} {\bibinfo {author} {\bibfnamefont {V.}~\bibnamefont
  {Milner}}, \bibinfo {author} {\bibfnamefont {J.~L.}\ \bibnamefont {Hanssen}},
  \bibinfo {author} {\bibfnamefont {W.~C.}\ \bibnamefont {Campbell}}, \ and\
  \bibinfo {author} {\bibfnamefont {M.~G.}\ \bibnamefont {Raizen}},\ }\href
  {\doibase 10.1103/PhysRevLett.86.1514} {\bibfield  {journal} {\bibinfo
  {journal} {Phys. Rev. Lett.}\ }\textbf {\bibinfo {volume} {86}},\ \bibinfo
  {pages} {1514} (\bibinfo {year} {2001})}\BibitemShut {NoStop}%
\bibitem [{\citenamefont {Chalony}\ \emph {et~al.}(2013)\citenamefont
  {Chalony}, \citenamefont {Barr\'e}, \citenamefont {Marcos}, \citenamefont
  {Olivetti},\ and\ \citenamefont {Wilkowski}}]{Chalony2013}%
  \BibitemOpen
  \bibfield  {author} {\bibinfo {author} {\bibfnamefont {M.}~\bibnamefont
  {Chalony}}, \bibinfo {author} {\bibfnamefont {J.}~\bibnamefont {Barr\'e}},
  \bibinfo {author} {\bibfnamefont {B.}~\bibnamefont {Marcos}}, \bibinfo
  {author} {\bibfnamefont {A.}~\bibnamefont {Olivetti}}, \ and\ \bibinfo
  {author} {\bibfnamefont {D.}~\bibnamefont {Wilkowski}},\ }\href {\doibase
  10.1103/PhysRevA.87.013401} {\bibfield  {journal} {\bibinfo  {journal} {Phys.
  Rev. A}\ }\textbf {\bibinfo {volume} {87}},\ \bibinfo {pages} {013401}
  (\bibinfo {year} {2013})}\BibitemShut {NoStop}%
\bibitem [{\citenamefont {Rybicki}(1971)}]{Rybicki1971}%
  \BibitemOpen
  \bibfield  {author} {\bibinfo {author} {\bibfnamefont {G.~B.}\ \bibnamefont
  {Rybicki}},\ }\href {\doibase 10.1007/BF00649195} {\bibfield  {journal}
  {\bibinfo  {journal} {Astrophysics and Space Science}\ }\textbf {\bibinfo
  {volume} {14}},\ \bibinfo {pages} {56} (\bibinfo {year} {1971})}\BibitemShut
  {NoStop}%
\bibitem [{\citenamefont {Yawn}\ and\ \citenamefont
  {Miller}(1997)}]{Miller1997}%
  \BibitemOpen
  \bibfield  {author} {\bibinfo {author} {\bibfnamefont {K.~R.}\ \bibnamefont
  {Yawn}}\ and\ \bibinfo {author} {\bibfnamefont {B.~N.}\ \bibnamefont
  {Miller}},\ }\href {\doibase 10.1103/PhysRevLett.79.3561} {\bibfield
  {journal} {\bibinfo  {journal} {Phys. Rev. Lett.}\ }\textbf {\bibinfo
  {volume} {79}},\ \bibinfo {pages} {3561} (\bibinfo {year}
  {1997})}\BibitemShut {NoStop}%
\bibitem [{\citenamefont {Miller}\ and\ \citenamefont
  {Youngkins}(1998)}]{Miller1998}%
  \BibitemOpen
  \bibfield  {author} {\bibinfo {author} {\bibfnamefont {B.~N.}\ \bibnamefont
  {Miller}}\ and\ \bibinfo {author} {\bibfnamefont {P.}~\bibnamefont
  {Youngkins}},\ }\href {\doibase 10.1103/PhysRevLett.81.4794} {\bibfield
  {journal} {\bibinfo  {journal} {Phys. Rev. Lett.}\ }\textbf {\bibinfo
  {volume} {81}},\ \bibinfo {pages} {4794} (\bibinfo {year}
  {1998})}\BibitemShut {NoStop}%
\bibitem [{\citenamefont {Lauritzen}\ \emph {et~al.}(2013)\citenamefont
  {Lauritzen}, \citenamefont {Gustainis},\ and\ \citenamefont
  {Mann}}]{Lauritzen2013}%
  \BibitemOpen
  \bibfield  {author} {\bibinfo {author} {\bibfnamefont {A.}~\bibnamefont
  {Lauritzen}}, \bibinfo {author} {\bibfnamefont {P.}~\bibnamefont
  {Gustainis}}, \ and\ \bibinfo {author} {\bibfnamefont {R.~B.}\ \bibnamefont
  {Mann}},\ }\href {\doibase http://dx.doi.org/10.1063/1.4815834} {\bibfield
  {journal} {\bibinfo  {journal} {Journal of Mathematical Physics}\ }\textbf
  {\bibinfo {volume} {54}},\ \bibinfo {eid} {072703} (\bibinfo {year} {2013}),\
  http://dx.doi.org/10.1063/1.4815834}\BibitemShut {NoStop}%
\bibitem [{\citenamefont {Kumar}\ and\ \citenamefont
  {Miller}(2014)}]{Kumar2014}%
  \BibitemOpen
  \bibfield  {author} {\bibinfo {author} {\bibfnamefont {P.}~\bibnamefont
  {Kumar}}\ and\ \bibinfo {author} {\bibfnamefont {B.~N.}\ \bibnamefont
  {Miller}},\ }\href {\doibase 10.1103/PhysRevE.90.062918} {\bibfield
  {journal} {\bibinfo  {journal} {Phys. Rev. E}\ }\textbf {\bibinfo {volume}
  {90}},\ \bibinfo {pages} {062918} (\bibinfo {year} {2014})}\BibitemShut
  {NoStop}%
\bibitem [{\citenamefont {Lehtihet}\ and\ \citenamefont
  {Miller}(1986)}]{Lehtihet1986}%
  \BibitemOpen
  \bibfield  {author} {\bibinfo {author} {\bibfnamefont {H.}~\bibnamefont
  {Lehtihet}}\ and\ \bibinfo {author} {\bibfnamefont {B.}~\bibnamefont
  {Miller}},\ }\href {\doibase http://dx.doi.org/10.1016/0167-2789(86)90080-1}
  {\bibfield  {journal} {\bibinfo  {journal} {Physica D: Nonlinear Phenomena}\
  }\textbf {\bibinfo {volume} {21}},\ \bibinfo {pages} {93 } (\bibinfo {year}
  {1986})}\BibitemShut {NoStop}%
\bibitem [{\citenamefont {Burnell}\ \emph {et~al.}(2003)\citenamefont
  {Burnell}, \citenamefont {Mann},\ and\ \citenamefont {Ohta}}]{Burnell2003}%
  \BibitemOpen
  \bibfield  {author} {\bibinfo {author} {\bibfnamefont {F.}~\bibnamefont
  {Burnell}}, \bibinfo {author} {\bibfnamefont {R.~B.}\ \bibnamefont {Mann}}, \
  and\ \bibinfo {author} {\bibfnamefont {T.}~\bibnamefont {Ohta}},\ }\href
  {\doibase 10.1103/PhysRevLett.90.134101} {\bibfield  {journal} {\bibinfo
  {journal} {Phys. Rev. Lett.}\ }\textbf {\bibinfo {volume} {90}},\ \bibinfo
  {pages} {134101} (\bibinfo {year} {2003})}\BibitemShut {NoStop}%
\bibitem [{\citenamefont {Burnell}\ \emph {et~al.}(2004)\citenamefont
  {Burnell}, \citenamefont {Malecki}, \citenamefont {Mann},\ and\ \citenamefont
  {Ohta}}]{Burnell2004}%
  \BibitemOpen
  \bibfield  {author} {\bibinfo {author} {\bibfnamefont {F.}~\bibnamefont
  {Burnell}}, \bibinfo {author} {\bibfnamefont {J.~J.}\ \bibnamefont
  {Malecki}}, \bibinfo {author} {\bibfnamefont {R.~B.}\ \bibnamefont {Mann}}, \
  and\ \bibinfo {author} {\bibfnamefont {T.}~\bibnamefont {Ohta}},\ }\href
  {\doibase 10.1103/PhysRevE.69.016214} {\bibfield  {journal} {\bibinfo
  {journal} {Phys. Rev. E}\ }\textbf {\bibinfo {volume} {69}},\ \bibinfo
  {pages} {016214} (\bibinfo {year} {2004})}\BibitemShut {NoStop}%
\bibitem [{\citenamefont {Springiel}\ \emph {et~al.}(2006)\citenamefont
  {Springiel}, \citenamefont {Frenk},\ and\ \citenamefont
  {White}}]{Springiel2006}%
  \BibitemOpen
  \bibfield  {author} {\bibinfo {author} {\bibfnamefont {V.}~\bibnamefont
  {Springiel}}, \bibinfo {author} {\bibfnamefont {C.~S.}\ \bibnamefont
  {Frenk}}, \ and\ \bibinfo {author} {\bibfnamefont {S.~D.~M.}\ \bibnamefont
  {White}},\ }\href@noop {} {\bibfield  {journal} {\bibinfo  {journal}
  {Nature}\ }\textbf {\bibinfo {volume} {440}},\ \bibinfo {pages} {1137}
  (\bibinfo {year} {2006})}\BibitemShut {NoStop}%
\bibitem [{\citenamefont {Bertschinger}(1998)}]{Bertschinger1998}%
  \BibitemOpen
  \bibfield  {author} {\bibinfo {author} {\bibfnamefont {E.}~\bibnamefont
  {Bertschinger}},\ }\href@noop {} {\bibfield  {journal} {\bibinfo  {journal}
  {Annual Review of Astronomy and Astrophysics}\ }\textbf {\bibinfo {volume}
  {36}},\ \bibinfo {pages} {599} (\bibinfo {year} {1998})}\BibitemShut
  {NoStop}%
\bibitem [{\citenamefont {Hockney}\ and\ \citenamefont
  {Eastwood}(1988)}]{hockney1988}%
  \BibitemOpen
  \bibfield  {author} {\bibinfo {author} {\bibfnamefont {R.~W.}\ \bibnamefont
  {Hockney}}\ and\ \bibinfo {author} {\bibfnamefont {J.~W.}\ \bibnamefont
  {Eastwood}},\ }\href {http://books.google.com/books?id=nTOFkmnCQuIC} {\emph
  {\bibinfo {title} {Computer Simulation Using Particles}}}\ (\bibinfo
  {publisher} {Taylor \& Francis},\ \bibinfo {year} {1988})\BibitemShut
  {NoStop}%
\bibitem [{\citenamefont {Hernquist}\ \emph {et~al.}(1991)\citenamefont
  {Hernquist}, \citenamefont {Bouchet},\ and\ \citenamefont
  {Suto}}]{Hernquist1991}%
  \BibitemOpen
  \bibfield  {author} {\bibinfo {author} {\bibfnamefont {L.}~\bibnamefont
  {Hernquist}}, \bibinfo {author} {\bibfnamefont {F.~R.}\ \bibnamefont
  {Bouchet}}, \ and\ \bibinfo {author} {\bibfnamefont {Y.}~\bibnamefont
  {Suto}},\ }\href {\doibase 10.1086/191530} {\bibfield  {journal} {\bibinfo
  {journal} {Astrophysical Journal Supplement}\ }\textbf {\bibinfo {volume}
  {75}},\ \bibinfo {pages} {231} (\bibinfo {year} {1991})}\BibitemShut
  {NoStop}%
\bibitem [{\citenamefont {Kunz}(1974)}]{Kunz1974}%
  \BibitemOpen
  \bibfield  {author} {\bibinfo {author} {\bibfnamefont {H.}~\bibnamefont
  {Kunz}},\ }\href {\doibase http://dx.doi.org/10.1016/0003-4916(74)90413-8}
  {\bibfield  {journal} {\bibinfo  {journal} {Annals of Physics}\ }\textbf
  {\bibinfo {volume} {85}},\ \bibinfo {pages} {303 } (\bibinfo {year}
  {1974})}\BibitemShut {NoStop}%
\bibitem [{\citenamefont {Schotte}\ and\ \citenamefont
  {Truong}(1980)}]{Schotte1980}%
  \BibitemOpen
  \bibfield  {author} {\bibinfo {author} {\bibfnamefont {K.~D.}\ \bibnamefont
  {Schotte}}\ and\ \bibinfo {author} {\bibfnamefont {T.~T.}\ \bibnamefont
  {Truong}},\ }\href {\doibase 10.1103/PhysRevA.22.2183} {\bibfield  {journal}
  {\bibinfo  {journal} {Phys. Rev. A}\ }\textbf {\bibinfo {volume} {22}},\
  \bibinfo {pages} {2183} (\bibinfo {year} {1980})}\BibitemShut {NoStop}%
\bibitem [{\citenamefont {Miller}\ and\ \citenamefont
  {Rouet}(2010)}]{miller2010}%
  \BibitemOpen
  \bibfield  {author} {\bibinfo {author} {\bibfnamefont {B.~N.}\ \bibnamefont
  {Miller}}\ and\ \bibinfo {author} {\bibfnamefont {J.-L.}\ \bibnamefont
  {Rouet}},\ }\href@noop {} {\bibfield  {journal} {\bibinfo  {journal}
  {Physical Review E}\ }\textbf {\bibinfo {volume} {82}},\ \bibinfo {pages}
  {066203} (\bibinfo {year} {2010})}\BibitemShut {NoStop}%
\bibitem [{\citenamefont {Arnol'd}\ and\ \citenamefont
  {Avez}(1968)}]{Arnold1968}%
  \BibitemOpen
  \bibfield  {author} {\bibinfo {author} {\bibfnamefont {V.~I.}\ \bibnamefont
  {Arnol'd}}\ and\ \bibinfo {author} {\bibfnamefont {A.}~\bibnamefont {Avez}},\
  }\href {http://cds.cern.ch/record/1987366} {\emph {\bibinfo {title} {{Ergodic
  problems of classical mechanics}}}},\ The mathematical physics monograph
  series\ (\bibinfo  {publisher} {W. A. Benjamin},\ \bibinfo {address} {New
  York, NY},\ \bibinfo {year} {1968})\BibitemShut {NoStop}%
\bibitem [{\citenamefont {Benettin}\ \emph {et~al.}(1979)\citenamefont
  {Benettin}, \citenamefont {Froeschle},\ and\ \citenamefont
  {Scheidecker}}]{Benettin1979}%
  \BibitemOpen
  \bibfield  {author} {\bibinfo {author} {\bibfnamefont {G.}~\bibnamefont
  {Benettin}}, \bibinfo {author} {\bibfnamefont {C.}~\bibnamefont {Froeschle}},
  \ and\ \bibinfo {author} {\bibfnamefont {J.~P.}\ \bibnamefont
  {Scheidecker}},\ }\href {\doibase 10.1103/PhysRevA.19.2454} {\bibfield
  {journal} {\bibinfo  {journal} {Phys. Rev. A}\ }\textbf {\bibinfo {volume}
  {19}},\ \bibinfo {pages} {2454} (\bibinfo {year} {1979})}\BibitemShut
  {NoStop}%
\end{thebibliography}%
\end{document}